\documentclass[epj]{webofc}
\usepackage[varg]{txfonts}   
\usepackage{natbib}

\woctitle{Hot Planets and Cool Stars}

\begin{document}

\title{Hot Moons and Cool Stars}

\author{Ren\'e Heller\inst{1}\fnsep\thanks{\email{rheller@aip.de}} \and
        Rory Barnes\inst{2,3}\fnsep\thanks{\email{rory@astro.washington.edu}}
}

\institute{Leibniz-Institut f\"ur Astrophysik Potsdam (AIP), An der Sternwarte 16, 14482 Potsdam, Germany 
\and
           University of Washington, Dept. of Astronomy, Seattle, WA 98195
\and
           Virtual Planetary Laboratory, USA
          }

\abstract{The exquisite photometric precision of the \textit{Kepler} space telescope now puts the detection of extrasolar moons at the horizon. Here, we firstly review observational and analytical techniques that have recently been proposed to find exomoons. Secondly, we discuss the prospects of characterizing potentially habitable extrasolar satellites. With moons being much more numerous than planets in the solar system and with most exoplanets found in the stellar habitable zone being gas giants, habitable moons could be as abundant as habitable planets. However, satellites orbiting planets in the habitable zones of cool stars will encounter strong tidal heating and likely appear as hot moons.}

\maketitle

\section{Introduction}

The advent of high-precision photometry from space with the \textit{CoRoT} and \textit{Kepler} telescopes has dramatically increased the number of confirmed and putative extrasolar planets. Beyond the sheer number of detections, smaller and smaller exoplanets were found with the today record being about $0.5$ Earth-radii \citep{2012ApJ...747..144M}. This achievement is of paramount importance for astrobiological investigations, as roughly Earth-sized planets may be inhabited -- provided many other requirements are met, of course.

Habitability of terrestrial planets can formally be defined as a planet's ability to allow for liquid surface water. The presence of liquid surface water will depend on the planet's distance to its host star, amongst others, with the adequate distance range spanning the stellar habitable zone, depending on the planet's atmospheric composition and surface pressure \citep{1993Icar..101..108K,2007A&A...476.1373S}. As of today, roughly $50$ extrasolar planet candidates have been confirmed in the \textit{Kepler} data, most of which are much bigger than Earth. These planets are likely to be gaseous and to resemble Neptune or Jupiter, rather than Earth. While they are not likely to be habitable, their moons might be.

Referred to exomoons, Martin Still, Director of the Kepler Guest Observer Office, said in his talk on the extended \textit{Kepler} project during this meeting: ``They are gonna come.'' So how can massive extrasolar moons be detected, provided they exist in the first place? And to which extent will they possibly be characterized?

\section{Detection methods}
\label{sec:detection}

\subsection{Transit timing and duration variations of the planet}

Two of the most promising techniques proposed for finding exomoons are transit timing variations (TTVs) and transit duration variations (TDVs) of the host planet. Combination of TTV and TDV measurements can provide information about a satellite's mass, its semi-major axis around the planet \citep{1999A&AS..134..553S,2007A&A...470..727S,2009MNRAS.392..181K}, and possibly about the inclination of the satellite's orbit with respect to the orbit around the star \citep{2009MNRAS.396.1797K}. The first dedicated hunt for exomoons in the \textit{Kepler} data is now underway \citep{2012ApJ...750..115K} and could possibly detect exomoons with masses down to 20\% the mass of Earth \citep{2009MNRAS.400..398K}. This corresponds to roughly $10$ times the mass of the two most massive moons in the solar system, Ganymede and Titan.

\subsection{Direct observations of the moon}

Observations of an exomoon transit itself \citep{2006A&A...450..395S,2011ApJ...743...97T,2011EPJWC..1101009L,2011MNRAS.416..689K} as well as planet-satellite mutual eclipses \citep{2007A&A...464.1133C,2012MNRAS.420.1630P} can provide information about the satellite's radius. Spectroscopic investigations of a moon's Rossiter-McLaughlin effect can yield information about its orbital geometry \citep{2010MNRAS.406.2038S,2012ApJ...758..111Z}, although relevant effects require accuracies in stellar radial velocity of the order of a few cm/s.

\begin{figure*}[t]
  \flushleft
  \scalebox{0.223}{\includegraphics{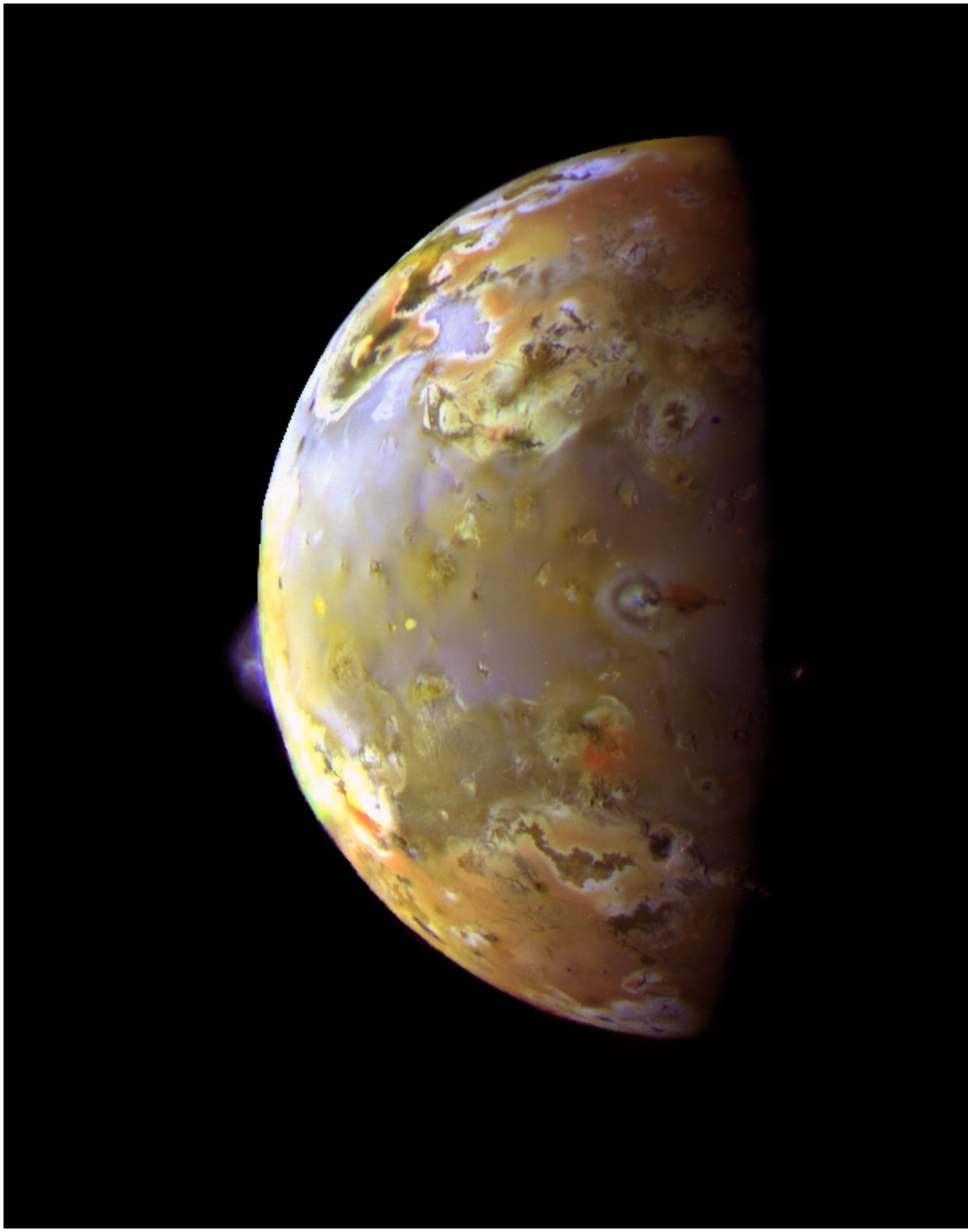}}
  \hspace{2.9cm}
  \scalebox{0.223}{\includegraphics{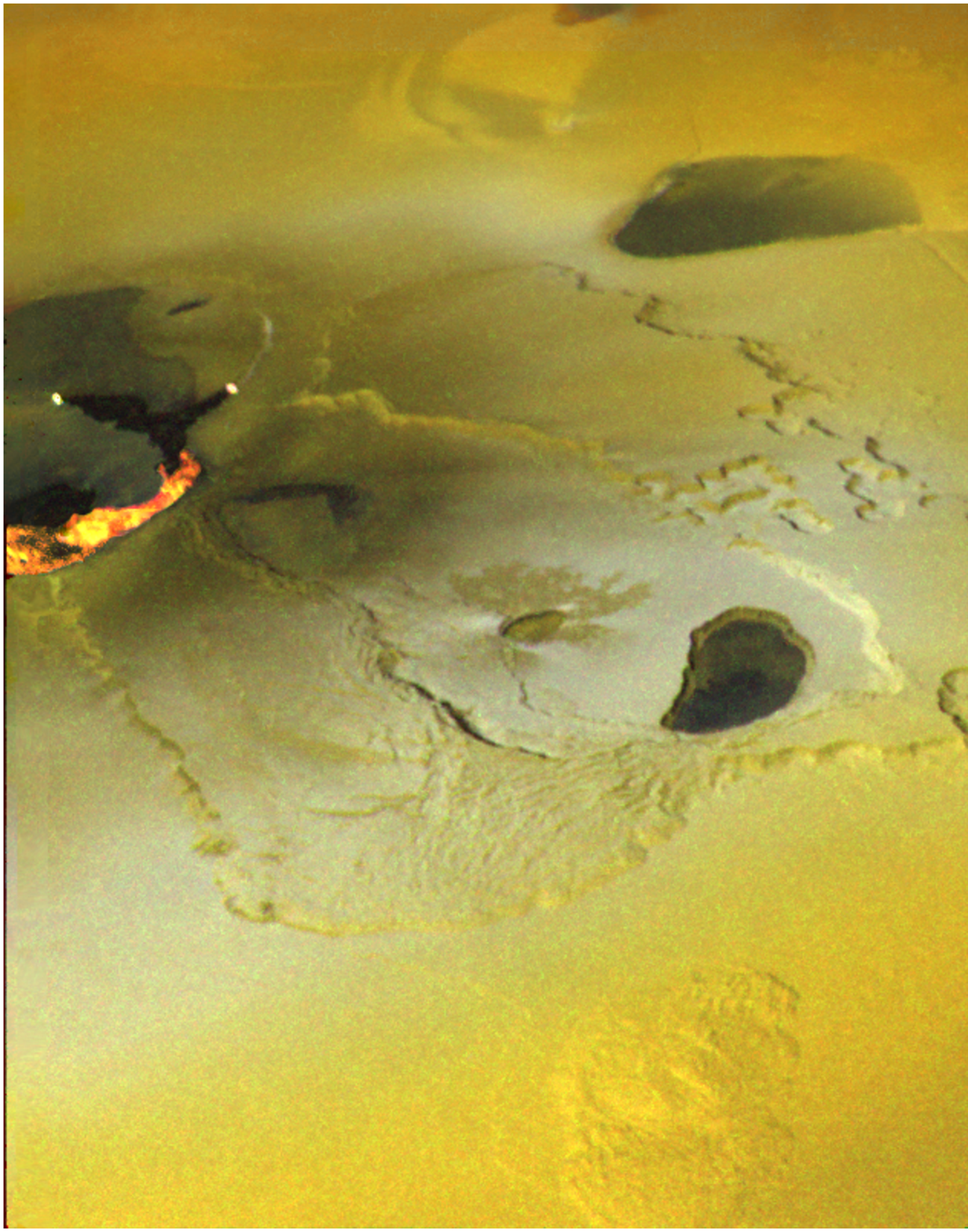}}
  \caption{The tectonically most active body in the solar system is Jupiter's moon Io. Its enhanced volcanism is driven by tidal dissipation inside the satellite. The two inlets in the left image, taken by the Galileo Orbiter, show a sulfuric plume over a volcanic depression named Pillan Patera (upper photograph) and another eruption called the Prometheus plume (lower photograph). The right image shows the strong orange infrared emission of flowing lava in Tvashtar Catena, a chain of calderas on Io. In extrasolar moons, tidal heating may be much stronger and even make them detectable via direct imaging. (Image credits: NASA/JPL)}
  \label{fig:Io}
\end{figure*}

Moons in the stellar habitable zone of low-mass stars must orbit their host planet very closely to remain gravitationally bound \citep{2012A&A...545L...8H,2012arXiv1210.5172H}. This will trigger enhanced tidal heating on those hypothetical moons and could make them uninhabitable. While a threat to life, enormous tidal heating in terrestrial moons about giant planets could be strong enough to make them detectable by direct imaging \citep{2012arXiv1209.4418P}. Tidal heating in moons has been observed in the solar system, with Jupiter's moon Io serving as the most prominent example (see Fig.~\ref{fig:Io}). As tidal heating in a satellite is proportional to the host planet's mass cubed, massive planets provide the most promising targets for direct imaging detections of tidally heated exomoons.

\section{Characterizing exomoons in the stellar habitable zone}

\begin{figure*}[t]
  \flushleft
  \scalebox{0.2155}{\includegraphics{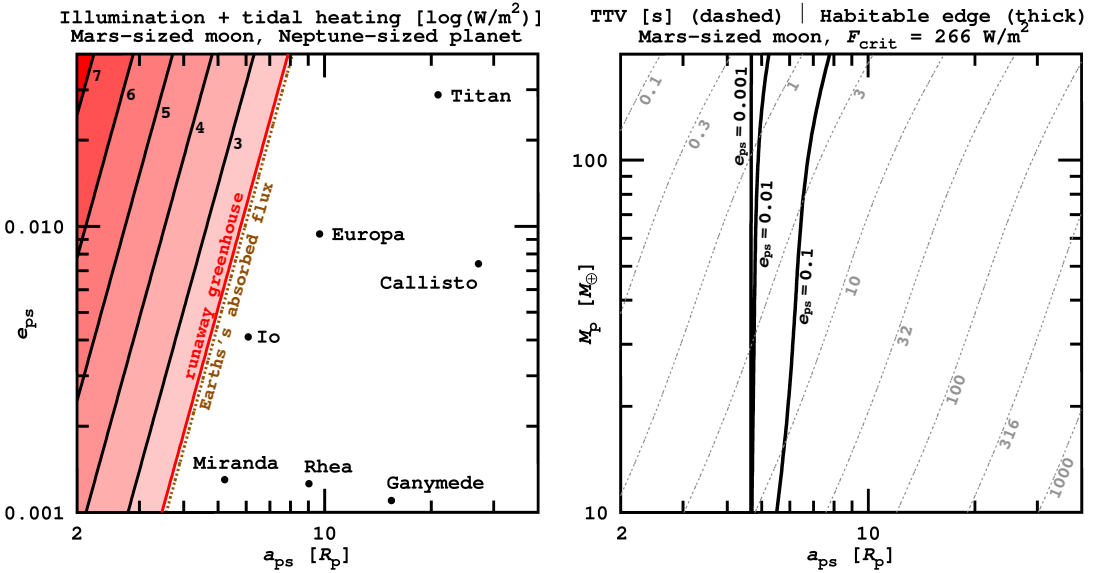}}
  \caption{\textit{Left:} Total top-of-the-atmosphere flux (in logarithmic units of W/m$^2$) of a Mars-sized moon about a Neptune-sized planet in the habitable zone of a $0.25\,M_\odot$ star. Tidal heating increases with decreasing semi-major axis $a_\mathrm{ps}$ (abscissa) and increasing eccentricity $e_\mathrm{ps}$ (ordinate). Some examples for orbital elements of solar system moons are indicated. \textit{Right:} Amplitude of the transit timing variation (dashed lines) for a Mass-sized moon about a range of host planets. Planetary masses are shown in Earth masses on the ordinate. The habitable edge \citep{2012arXiv1209.5323H} is indicated for three different orbital eccentricities $e_\mathrm{ps}$ of the satellite: $0.1$, $0.01$, and $0.001$.}
  \label{fig:heat_TTV}
\end{figure*}

A massive moon in the stellar habitable zone can be subject to strong tidal heating and hence be uninhabitable. To estimate its habitability, we have set up a model that includes stellar and planetary illumination as well as tidal heating \citep{2012arXiv1209.5323H}. If their sum is greater than the critical flux for the moon to be subject to a runaway greenhouse effect \citep{2010ppc..book.....P}, the moon will lose all its water and become uninhabitable. In the left panel of Fig.~\ref{fig:heat_TTV} we show contours of the orbit-averaged illumination plus tidal heat flux of a hypothetical Mars-sized moon orbiting a Neptune-sized planet in the habitable zone of a $0.25\,M_\odot$ star. Abscissa indicates the planet-satellite semi-major axis in planetary radii, ordinate shows orbital eccentricity. In the reddish regions this prototype satellite will be desiccated and uninhabitable.

Given sufficiently long observational coverage and high-accuracy data, the techniques described in Sect.~\ref{sec:detection} make it possible to detect and considerably characterize a sub-Earth-sized moon orbiting a giant planet. In the right panel of Fig.~\ref{fig:heat_TTV} we show the TTV amplitudes of a Mars-sized moon orbiting a range of host planets. It is assumed that the moon's orbit is circular and that both the circumstellar and the circum-planetary orbit are seen edge-on from Earth \citep{2012ApJ...750..115K}.

To determine a satellite's mass and orbit via TTV and TDV, many transits of the host planet need to be observed. Thus, to find moons about planets in the stellar habitable zone within the \textit{Kepler} duty cycle of 7 years, one might be tempted to conclude that they can preferably be detected in cool star systems (i.e. around M dwarfs), because their habitable zones are close-by where a planet performs potentially many transits in a given time span. However, the amplitude of a planet's TTV is smaller in cool star systems, given a fixed semi-major axis \citep{2006A&A...450..395S}, and the lack of M dwarfs in the \textit{Kepler} sample \citep{2012arXiv1202.5852B} further decreases the chance of finding habitable moons in cool star systems.

Beyond that, moons of planets in the habitable zones of cool stars might not be habitable in the first place \citep{2012A&A...545L...8H}. The planet's sphere of gravitational dominance, i.e. its Hill sphere, is relatively small due to the close star. Hence, any moon would have to follow a very tight orbit about the planet. Moreover, the close star will force the satellite's orbit to be non-circular. Both aspects, small orbital distance and an eccentric orbit, will cause any Earth-sized moon of a massive gaseous planet to experience enormous tidal heating. Ultimately, stellar irradiation and tidal heating will sum up to a top-of-the-atmosphere energy flux that exceeds the critical flux for the initiation of the runaway greenhouse effect \citep{2012arXiv1209.5323H}. Moons of planets in the habitable zones of cool stars will thus be hot rather than habitable.

To characterize potentially habitable moons, that is to say, moons in the habitable zones of K and G type stars, using the TTV and TDV techniques will take about a decade of observations, at least. With the \textit{Kepler} mission being scheduled for a total mission cycle of 7 years, such a detection might just be at the edge of what is possible \citep{2006A&A...450..395S,2009MNRAS.400..398K,2012ApJ...750..115K}.

\bibliography{2012-7-HellerBarnes-Exomoons_ROPACS_Proceedings}

\begin{thebibliography}{23}

\bibitem{2012ApJ...747..144M}
P.S. {Muirhead}, J.A. {Johnson}, K.~{Apps}, J.A. {Carter}, T.D. {Morton}, D.C.
  {Fabrycky}, J.S. {Pineda}, M.~{Bottom}, B.~{Rojas-Ayala}, E.~{Schlawin}
  et~al., \textit{ApJ} \textbf{747}, 144 (2012), \texttt{1201.2189}

\bibitem{1993Icar..101..108K}
J.F. {Kasting}, D.P. {Whitmire}, R.T. {Reynolds}, \textit{Icarus} \textbf{101},
  108 (1993)

\bibitem{2007A&A...476.1373S}
F.~{Selsis}, J.F. {Kasting}, B.~{Levrard}, J.~{Paillet}, I.~{Ribas},
  X.~{Delfosse}, \textit{A\&A} \textbf{476}, 1373 (2007), \texttt{0710.5294}

\bibitem{1999A&AS..134..553S}
P.~{Sartoretti}, J.~{Schneider}, \textit{A\&AS} \textbf{134}, 553 (1999)

\bibitem{2007A&A...470..727S}
A.~{Simon}, K.~{Szatm{\'a}ry}, G.M. {Szab{\'o}}, \textit{A\&A} \textbf{470},
  727 (2007), \texttt{0705.1046}

\bibitem{2009MNRAS.392..181K}
D.M. {Kipping}, \textit{MNRAS} \textbf{392}, 181 (2009), \texttt{0810.2243}

\bibitem{2009MNRAS.396.1797K}
D.M. {Kipping}, \textit{MNRAS} \textbf{396}, 1797 (2009), \texttt{0904.2565}

\bibitem{2012ApJ...750..115K}
D.M. {Kipping}, G.{\'A}. {Bakos}, L.~{Buchhave}, D.~{Nesvorn{\'y}},
  A.~{Schmitt}, \textit{ApJ} \textbf{750}, 115 (2012), \texttt{1201.0752}

\bibitem{2009MNRAS.400..398K}
D.M. {Kipping}, S.J. {Fossey}, G.~{Campanella}, \textit{MNRAS} \textbf{400},
  398 (2009), \texttt{0907.3909}

\bibitem{2006A&A...450..395S}
G.M. {Szab{\'o}}, K.~{Szatm{\'a}ry}, Z.~{Div{\'e}ki}, A.~{Simon}, \textit{A\&A}
  \textbf{450}, 395 (2006), \texttt{0601186}

\bibitem{2011ApJ...743...97T}
L.R.M. {Tusnski}, A.~{Valio}, \textit{ApJ} \textbf{743}, 97 (2011),
  \texttt{1111.5599}

\bibitem{2011EPJWC..1101009L}
K.~{Lewis}, in \emph{European Physical Journal Web of Conferences} (2011),
  Vol.~11, p. 1009

\bibitem{2011MNRAS.416..689K}
D.M. {Kipping}, \textit{MNRAS} \textbf{416}, 689 (2011), \texttt{1105.3499}

\bibitem{2007A&A...464.1133C}
J.~{Cabrera}, J.~{Schneider}, \textit{A\&A} \textbf{464}, 1133 (2007),
  \texttt{0703609}

\bibitem{2012MNRAS.420.1630P}
A.~{P{\'a}l}, \textit{MNRAS} \textbf{420}, 1630 (2012), \texttt{1111.1741}

\bibitem{2010MNRAS.406.2038S}
A.E. {Simon}, G.M. {Szab{\'o}}, K.~{Szatm{\'a}ry}, L.L. {Kiss}, \textit{MNRAS}
  \textbf{406}, 2038 (2010)

\bibitem{2012ApJ...758..111Z}
Q.~{Zhuang}, X.~{Gao}, Q.~{Yu}, \textit{ApJ} \textbf{758}, 111 (2012),
  \texttt{1207.6966}

\bibitem{2012A&A...545L...8H}
R.~{Heller}, \textit{A\&A} \textbf{545}, L8 (2012), \texttt{1209.0050}

\bibitem{2012arXiv1210.5172H}
R.~{Heller}, R.~{Barnes}, ArXiv e-prints  (2012), \texttt{1210.5172}

\bibitem{2012arXiv1209.4418P}
M.A. {Peters}, E.L. {Turner}, ArXiv e-prints  (2012), \texttt{1209.4418}

\bibitem{2012arXiv1209.5323H}
R.~{Heller}, R.~{Barnes}, \textit{Astrobiology} (in press)  (2013),
  \texttt{1209.5323}

\bibitem{2010ppc..book.....P}
R.T. {Pierrehumbert}, \emph{{Principles of Planetary Climate}} (2010)

\bibitem{2012arXiv1202.5852B}
N.M. {Batalha}, J.F. {Rowe}, S.T. {Bryson}, T.~{Barclay}, C.J. {Burke}, D.A.
  {Caldwell}, J.L. {Christiansen}, F.~{Mullally}, S.E. {Thompson}, T.M. {Brown}
  et~al., ArXiv e-prints  (2012), \texttt{1202.5852}

\end{thebibliography}

\end{document}